\documentclass[12pt,preprint]{aastex}

\shorttitle{VLA and VLBA Observations of the $z=6.12$ QSO J1427+3312}
\shortauthors{Momjian et al.}

\begin{document}

\title{VLA and VLBA Observations of the Highest Redshift Radio-Loud QSO J1427+3312 at $z = 6.12$}

\author{Emmanuel Momjian\altaffilmark{1}}
\affil{NAIC, Arecibo Observatory, HC 3, Box 53995, Arecibo, PR 00612}
\email{emomjian@naic.edu}

\author{Christopher L. Carilli}
\affil{National Radio Astronomy Observatory, P. O. Box O, Socorro, NM, 87801}
\email{ccarilli@nrao.edu}

\author{Ian D. McGreer}
\affil{Department of Astronomy, Columbia University, 550 West 120th Street, New York, NY 10027}
\email{mcgreer@astro.columbia.edu}

\altaffiltext{1}{now at: National Radio Astronomy Observatory, P. O. Box O, Socorro, NM, 87801, emomjian@nrao.edu} 
\begin{abstract}

We present 8.4~GHz VLA A-array and 1.4~GHz VLBA results on the radio continuum
emission from the highest redshift radio-loud quasar known to date, the $z=6.12$ QSO J1427+3312.
The VLA observations show an unresolved steep spectrum source with a flux density of
$250 \pm 20$~${\mu}$Jy at 8.4~GHz and a spectral index value of $\alpha^{8.4}_{1.4}=-1.1$.
The 1.4~GHz VLBA images reveal several continuum components with a total flux density of 
$1.778 \pm 0.109$~mJy, which is consistent with the flux density measured with the VLA at 1.4~GHz.
Each of these components is resolved with sizes of a few milliarcseconds, and intrinsic brightness
temperatures on the order of $10^7$ to $10^8$~K. The physical characteristics as revealed in these
observations suggest that this QSO may be a Compact Symmetric Object, with the two dominant components 
seen with the VLBA, which are separated by 31 mas (174 pc) and have intrinsic sizes of $\sim 22-34$pc,
being the two radio lobes that
are confined by the dense ISM. If indeed a CSO, then the estimated kinematic age of this radio AGN
is only $10^3$ yr.

\end{abstract}


\section{INTRODUCTION}

Optical surveys such as the Sloan Digital Sky Survey (SDSS; \citet{YOR00})
and the Digitized Palomar Sky Survey \citep{DJO99} have revealed large samples
of quasi-stellar objects out to $z \sim 6$. Studies by \citet{FAN02,FCK06}
have shown that at such a high redshift we are approaching the epoch of
reionization, the edge of the ``dark ages'', when the first stars and massive
black holes were formed.
Eddington limit arguments suggests that
the supermassive black holes at the center of these QSOs are on the order of
$10^9~M{_\odot}$. If the correlation between bulge and black hole masses \citep{GEB00,FEME00,KU07}
also holds at these high redshifts, then these sources have associated spheroids
with masses on the order of $\sim 10^{12}~M_\odot$. It is challenging to explain the formation
of such massive structures in relatively short timescales {($<$~1~Gyr; e.g. \citet{LI07})}.
\citet{WL02} estimate that almost one third of known quasars at $z\sim6$
ought to be lensed by galaxies along the line of sight. If these quasars are indeed
gravitationally lensed, the estimated masses of their associated spheroids could be smaller 
by up to an order of magnitude; this would allow a less efficient assembly process. However,
\citet{KU07} derived black hole masses in a sample of $z\sim6$ quasars using
the widths of the Mg {\footnotesize II} and C {\footnotesize IV} lines.
The resulting mass values were comparable and in agreement with those derived by assuming 
that the quasars emit at their Eddington luminosity, implying that the
quasars are not likely to be strongly lensed \citep{KU07}. 

High resolution radio observations of high-redshift radio-loud quasars
can be used to test for strong gravitational lensing by looking for
multiple imaging on scales from tens of milliarcseconds (mas) to
arcseconds. Also, Very Long Baseline Interferometry (VLBI) observations of
core-jet radio sources in
quasars over a large range in redshift have been used to constrain the
cosmic geometry, under the assumption that such sources are (roughly)
`standard rulers' \citep{GKF99}. In general, the high resolution of the VLBI
observations permit a more detailed look at the physical structures in the most
distant cosmic sources on scales unreachable with any other technique (mas).

To date, several radio-quiet and radio-loud quasars with redshifts $z > 4$
have been imaged using
VLBI \citep{FRE97,FRE03,MOM04,MOM05,MOM07}. Of these, only two
were radio-loud quasars at redshifts $z > 5$, namely J0913$+$5919 at
$z=5.11$ \citep{MOM04}, and J0836$+$0054
at $z=5.82$ \citep{FRE03}.
Furthermore, and until recently, all known $z > 6$ QSOs were radio-quiet. However,
\citet{MCG06} reported the discovery of the first radio-loud QSO at a
redshift greater than 6, namely the QSO J1427+3312 at $z=6.12$, by
matching VLA's FIRST survey \citep{BWH95} with the NOAO Deep
Wide Field Survey (NDWFS; \citet{JD99}) and the FLAMINGOS
EXtragalactic Survey (FLAMEX; \citet{ELS06}). 

In this paper we present 8.4~GHz Very Large Array (VLA) and 1.4~GHz Very Long Baseline
Array (VLBA)
observations of the highest redshift radio-loud quasar known to date, J1427+3312 at $z=6.12$.
At the redshift of the source, observing frequencies of 1.4 and 8.4~GHz correspond to rest frequencies of 10 and 60~GHz, respectively.

The source J1427+3312 is classified as a Broad Absorption Line Quasar
(BALQSO; \citet{MCG06}), and has a radio flux density of
1.73~mJy at 1.4 GHz \citep{BWH95}. 
This QSO is not detected in the {\it Chandra} XBo\"{o}tes survey \citep{MUR05}, implying
an upper limit in the 0.5-7 KeV band of $4 \times 10^{-15}$~erg~cm$^{-1}$~s$^{-1}$.

Spectra obtained with the Keck II telescope reveal two strong Mg II absorption systems, one
at $z=2.1804$ and another at $z=2.1997$ \citep{MCG06}. The
presence of these absorption systems in the line-of-sight raises the
possibility that the QSO is being gravitationally lensed.

Throughout this paper, we assume a flat cosmological model with
$\Omega_{m}=0.3$, $\Omega_\Lambda=0.7$, and
${H_{0}=71}$~km~s$^{-1}$~Mpc$^{-1}$. In this model, at the distance of J1427+3312,
1~mas corresponds to 5.6~pc.

\section{OBSERVATIONS AND DATA REDUCTION}

\subsection{VLA Observations}
The QSO J1427+3312 was observed with the Very Large Array (VLA) of the
NRAO\footnote{The National Radio Astronomy
Observatory is a facility of the National Science Foundation operated under
cooperative agreement by Associated Universities, Inc.} in A configuration on 
2007 July 9. The frequency of the observations was 8.4~GHz and the total bandwidth was
100~MHz in both right and left-hand circular polarizations.
The source 3C286 was used as the primary flux calibrator, and J1416+3444 as the phase calibrator.
The total time was 6 hours with 23 antennas participating in the observations. 
The VLA data was reduced using standard Astronomical Image
Processing System (AIPS) routines. 
Table~1 summarizes the parameters of the VLA observations.

\subsection{VLBA Observations}
The VLBI observations of J1427+3312 were carried out at 1.4~GHz on 2007 June 11 and 12,
using the Very Long Baseline Array (VLBA)
of the NRAO. Eight adjacent 8~MHz
baseband channel pairs were used in the observations, both with right and
left-hand circular polarizations, and sampled at two bits. The data were
correlated at the VLBA correlator in Socorro, NM, with 2~s correlator
integration time. The total observing time was 12~hr. 
Table~2 summarizes the parameters of the VLBA observations.

The VLBA observations employed nodding-style phase referencing, using the calibrator
J1422+3223  ($S_{\rm 1.4~GHz}=0.4$~Jy), with a cycle time of 4~min, 3~min on the
target source and 1~min
on the calibrator. The angular separation between the target source and the phase
calibrator is $1.4^{\circ}$.
A number of test cycles were also included to monitor the
coherence of the phase referencing. These tests involved switching between two
calibrators, the phase calibrator J1422+3223 and the phase-check calibrator
J1416+3444 ($S_{\rm 1.4~GHz}=1.9$~Jy), using a similar cycle time to that used for
the target source. The angular separation between the phase calibrator and the
phase-check calibrator is $2.7^{\circ}$.

The accuracy of the phase calibrator position is important in phase-referencing
observations \citep {WAL99}, as this determines the accuracy of the absolute
position of the target source and any associated components. Phase referencing,
as used here, is known to preserve absolute astrometric positions to better
than $\pm 0\rlap{.}^{''}01$ \citep{FOM99}.

Data reduction and analysis were performed using AIPS of the NRAO.
After applying {\it a priori} flagging, amplitude
calibration was performed using measurements of the antenna gain and system
temperature for each station. Ionospheric corrections were applied using the
AIPS task ``TECOR''. The phase calibrator J1422+3223 was self-calibrated in
both phase and amplitude and imaged in an iterative cycle.

Images of the phase-check calibrator, J1416+3444, were deconvolved using two
different approaches: (a) by applying the phase and the amplitude
self-calibration solutions of the phase reference source J1422+3223
(Figure~1{\it{a}}), and (b) by self calibrating J1416+3444 itself, in both
phase and amplitude (Figure~1{\it{b}}). The peak surface brightness ratio of
the final images from the two approaches gives a measure of the effect of
residual phase errors after phase referencing, i.e., `the coherence' due to
phase referencing \citep{MV95}. At all times, the coherence was found to be better than
99\%.

The self-calibration solutions of the phase calibrator, J1422+3223 (Fig.~2), were
applied on the target source, J1427+3312, which was then deconvolved and imaged.

\section{RESULTS \& ANALYSIS}

Figure~3 is a uniformly weighted image of the $z=6.12$ QSO J1427+3312 obtained with the
VLA A-array at 8.4~GHz and 0$\rlap{.}^{''}$19 (1.1~kpc) resolution. The
detected radio source is unresolved and the measured total flux density is $250 \pm 20$~${\mu}$Jy.
No other continuum features were found on scales of several arcseconds.

Comparing the 8.4~GHz flux density with previous 1.4 GHz measurements \citep{BWH95,CP99}
shows that this QSO is a steep spectrum source with a spectral index of $\alpha^{8.4}_{1.4}=-1.1$.

Figure~4 is a naturally weighted image of the target source at the full resolution of the
VLBA, which is 
$11.6 \times 7.8$~mas ($65 \times 42$~pc, PA=$2^{\circ}$). The rms noise level in this image is 
$28~\mu$Jy~beam$^{-1}$.

Table~3 lists the Gaussian fitting parameters of the compact
continuum sources seen in the VLBA image (Fig.~4) derived using the
AIPS task JMFIT. Gaussian components provide a convenient parameterization of source
structure even if they do not necessarily represent discrete physical structures.
Columns (2) and (3) are the coordinates of the sources, and column (4)
is the relative positions of the these sources with respect to the strongest component.
Column (5) lists the surface
brightnesses of these sources, and column (6) their
integrated flux densities. Column (7) gives the nominal deconvolved
sizes of the Gaussian components at FWHM as given by JMFIT, and
column 8 lists the position angles of the fitted Gaussians.
The corresponding intrinsic brightness temperatures of these compact sources are on
the order of $10^7$ to $10^8$~K, and are listed in column 9.

Based on the VLBA results, the source is composed of two dominant
structures separated by $\sim 31$~mas. The stronger of these is consistent with two
Gaussians (components 1 and 3 in Table~3), while the second dominant source is represented by one
Gaussian (component 2 in Table~3). Including a possible faint component to the
east (component 4 in Table~3), gives a total flux density of $1.778 \pm 
0.109$~mJy. This value is consistent with the $1.73 \pm 0.13$~mJy obtained with the VLA FIRST
survey \citep{BWH95}, and with the $1.816 \pm 0.021$~mJy obtained with the VLA ELAIS survey
\citep{CP99}. 
The 1.4 GHz flux densities measured with the VLA in 1995 and
1997 \citep{BWH95,CP99} and the VLBA in 2007 (this paper) are equal to better than 5\%,
implying that this source is not highly variable on time scales of years.

We have also synthesized larger images ($2'' \times 2''$) using the VLBA and found no
other radio components at $\ge 5\sigma$ level ($140~\mu$Jy~beam$^{-1}$) in the field
other than those seen in Figure~4 and listed in Table~3.

\section{DISCUSSION}

We have detected the $z=6.12$ QSO J1427+3312 at 8.4~GHz with the VLA A-array and at 1.4~GHz
with the VLBA. The source is unresolved as seen in the 8.4~GHz VLA results, and has a steep
spectrum with a spectral index value of $\alpha^{8.4}_{1.4}=-1.1$.

At mas resolution, the VLBA
observations show that this QSO is comprised of two dominant continuum components
separated by 31 mas (174 pc; Table~3) with a flux density ratio of $\sim 3:1$.
The Gaussian fitting suggests that both components are resolved with sizes of $\sim 4-6$~mas
($22-34$~pc).

The physical properties observed in this source suggest that this high-$z$ QSO could be
a Compact Symmetric Object (CSO) with two distinct, steep spectra, radio lobes that are confined by
a dense ISM in the host galaxy \citep{CO02}. 
CSOs are radio sources that have sizes on scales of 1~pc to 1~kpc, and are
thought to be very young ($\leq 10^4$~yr; \citet{RTXPWP96,OC98}). Moreover, because CSOs are highly confined
sources, they are thought to have higher
conversion efficiency of jet kinetic energy to radio luminosity.

For J1427+3312, the radio lobes would be the two dominant components seen in the 1.4~GHz
VLBA results, which are separated by 174~pc.
In this model, the core, which would have an inverted or flat spectrum \citep{TRP96},
would be faint and below our detection threshold. In CSOs, any radio component that is
coincident with the central engine is comparatively weak, with flux densities of only a few
percent of the total flux density \citep{RTXPWP96}.

Classifying this source as a CSO, where the radio lobes are confined by a dense ambient medium,
is also consistent with its being a BALQSO \citep{MCG06}. This is in agreement with recent X-ray observations
which suggest that the presence of massive, highly ionized, and high-velocity outflows in BALQSOs may be providing
significant feedback to the surrounding gas \citep{CH07}.
In general, the higher average gas densities expected in the earliest galaxies might lead
to a higher fraction of CSO sources.

Multi-epoch and multi-frequency VLBI observations carried out by \citet{TMPR00} on three
CSOs have shown typical advance speeds of $\sim 0.3c$. Adopting this value for J1427+3312,
and assuming equal speeds for the two radio lobes, we derive a kinematic age of $\sim 10^3$~yr,
indicating that the $z=6.12$ QSO is a very young radio source. 

With overwhelming majority of CSOs ($\geq$ 80\%) showing HI 21~cm absorption lines with optical depth 
levels between 4\% and 40\% and line widths ranging between about 50 and 500~km~s$^{-1}$
\citep{PECK00}, the $z=6.12$ QSO J1427+3312, which is believed to be near the Epoch of Reionization,
would be an excellent candidate for HI absorption experiments to detect the
neutral IGM in its host galaxy \citep{FL02}.
Such a search is currently underway using the
Giant Meterwave Radio Telescope (GMRT). Knowledge of source
structure, as presented herein, is critical for both identifying
potential candidates for HI 21~cm absorption searches and for
subsequent interpretation of the results.

The high resolution radio imaging presented here sets constraints on the hypothesis that
J1427+3312 has undergone strong gravitational lensing resulting in multiple images.
The VLA 8.4~GHz imaging rules out multiple images separated by $ > 0\rlap{.}^{''}2$ and with a 
flux ratio $< 8:1$.  The VLBA 1.4~GHz imaging sets similar constraints down to the mas scale.
The expected lensing system for a $z \sim 6$ quasar is a singular isothermal ellipsoid (SIE),
with image pairs having separations $\lesssim 1''$ and flux ratios $< 10:1$
\citep{TOG84}. The typical lens system
is thus ruled out, and only a high magnification or small separation lensing
event is allowed by the images obtained from these radio observations.

On the other hand, the possible faint component in the VLBA image to the east (component 4)
has a separation of $\sim 0\rlap{.}^{''}1$
from the strongest component and a flux ratio of $\sim 7:1$. While it is highly unlikely
that an SIE lens would result in such a small image separation, lensing by a spiral galaxy might.
However, spirals are expected to contribute only 10\%$-$20\% of gravitational lenses
\citep{TOG84,KK98}, and $< 5\%$ of spiral lenses would produce image separations
$\sim 0\rlap{.}^{''}1$ \citep{BL98}. Thus the a priori likelihood that components
4 and 1 represent a lensed image pair is quite low. Future sensitive VLBI observations
at multiple frequencies are needed to measure the spectral energy distribution of the
various components seen in the VLBA image and determine whether they have identical
physical characteristics.

We also cannot rule out that the continuum features seen in this QSO are core-jet
structures. The above mentioned sensitive VLBI observations will be able to address this
possibility as well. Furthermore, multi epoch VLBI observations can lead to determine
the proper motion of the components seen in J1427+3312, and provide a more accurate
estimate of the advance speeds of the radio lobes and the kinematic age of the source,
assuming that this $z=6.12$ is indeed a CSO, or the proper motion of the jet with
 respect to the core, if the continuum VLBA results were of core-jet structures.

\section{ACKNOWLEDGMENTS}

The Arecibo Observatory is part
of the National Astronomy and Ionosphere Center, which is operated by Cornell
University under a cooperative agreement with the National Science Foundation.
C.~L.~C. acknowledges support from the Max-Planck Society
and the Alexander von Humboldt Foundation through the Max-Planck
Forshungspreise 2005.

\begin{figure}
\epsscale{0.7}
\plotone{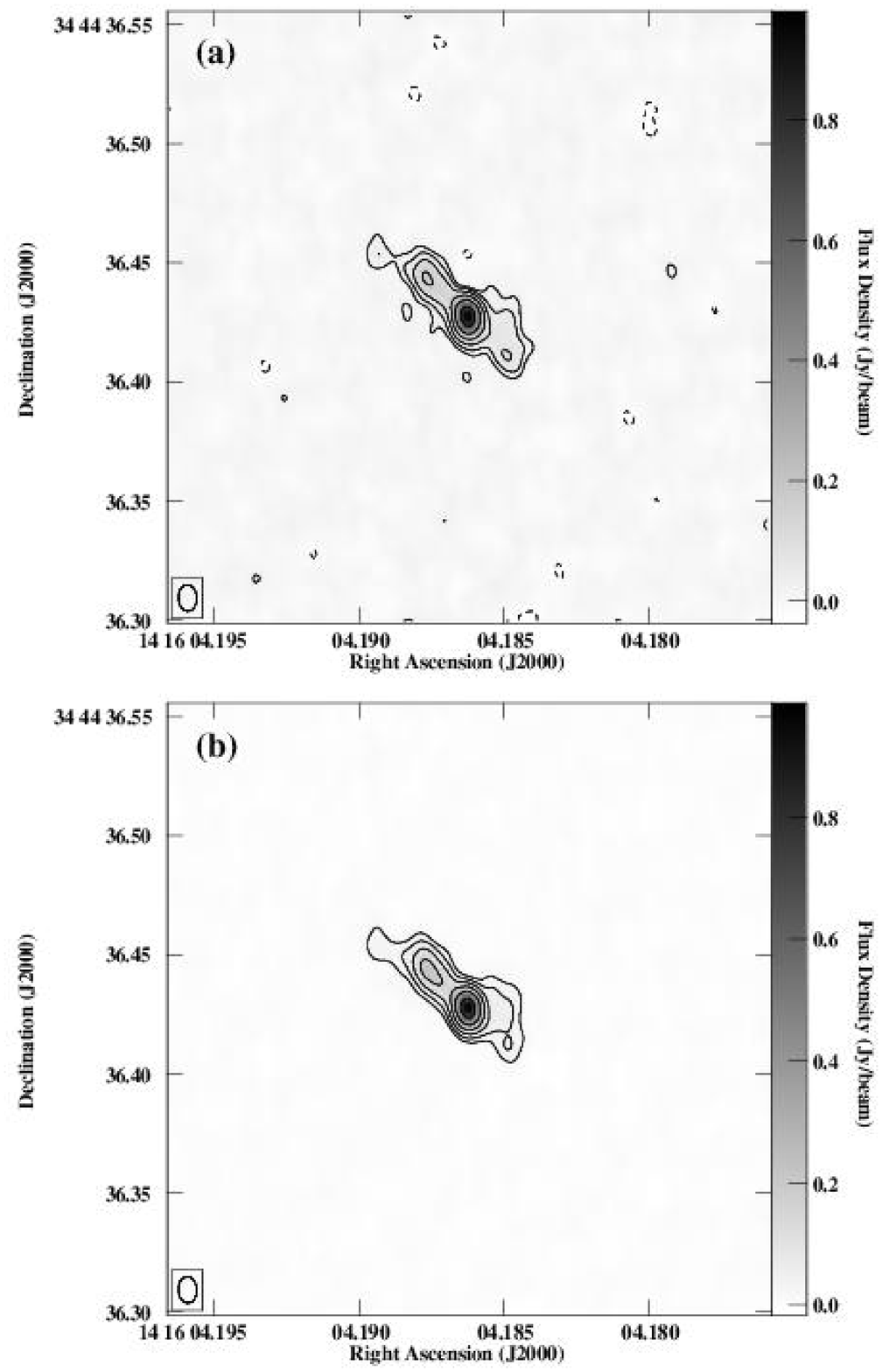}
\figcaption{VLBA continuum images of the phase-check
calibrator J1416+3444 at 1.4~GHz: a) obtained by applying the phase
and the amplitude self-calibration solutions of the phase reference
source J1422+3223, b) obtained by self calibrating J1416+3444
itself, in both phase and amplitude. The restoring beam size in both
images is $10.9 \times 7.8$~mas in position angle $3^{\circ}$. The
contour levels are at $-3$, 3, 6, 12, $\ldots$, 92  times the rms
noise level in the phase-referenced image (a), which is
7.1~mJy~beam$^{-1}$. The gray-scale range is indicated by the step
wedge at the right side of each image.
\label{f1}}
\end{figure}

\begin{figure}
\epsscale{1}
\plotone{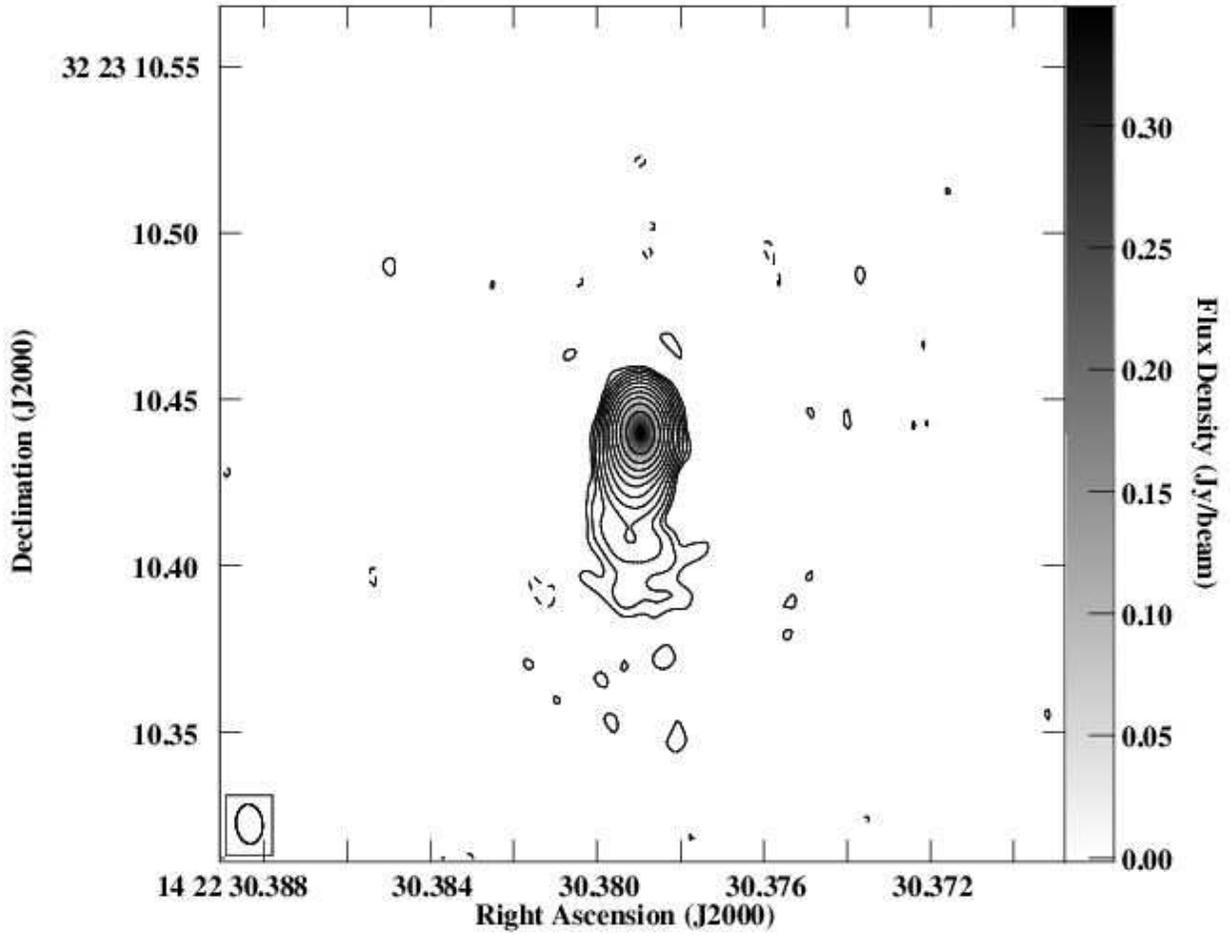}
\figcaption{VLBA continuum image of the phase
calibrator J1422+3223 at 1.4~GHz.
The restoring beam size is $11.9 \times 8.0$~mas in position angle $5^{\circ}$. The
contour levels are at $-3$, 3, 6, 12, $\ldots$, 6144 times the rms
noise level, which is
54~$\mu$Jy~beam$^{-1}$. The gray-scale range is indicated by the step
wedge at the right side of the image.
\label{f2}}
\end{figure}

\begin{figure}
\epsscale{1}
\plotone{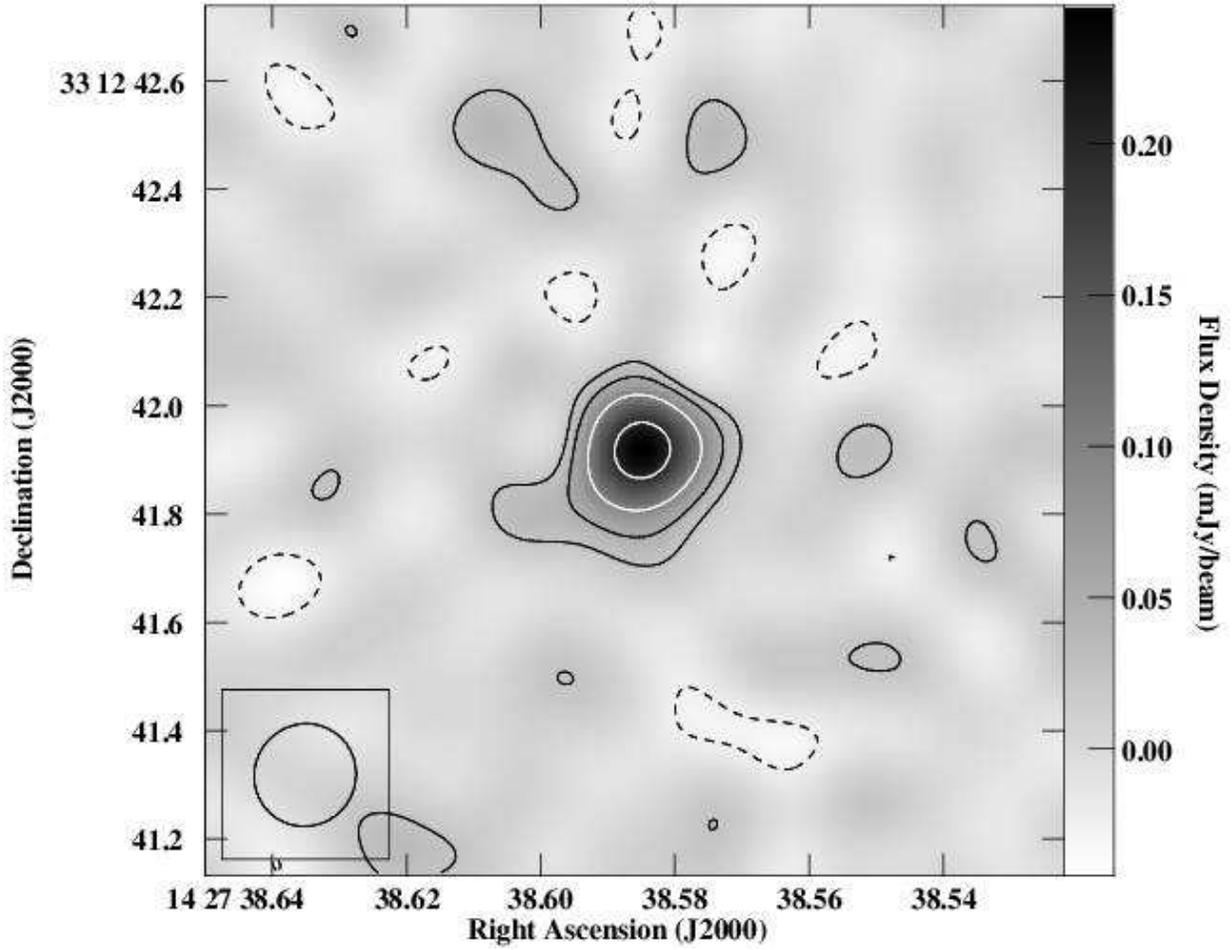}
\figcaption{Uniformly weighted VLA A-array continuum image of the $z=6.12$ QSO J1427+3312
at 8.4~GHz. 
The restoring beam size is $192 \times 188$~mas in position angle $-51^{\circ}$.
The contour levels are at $-2$, 2, 4, 8, 16  times the rms
noise level, which is 12.3~$\mu$Jy~beam$^{-1}$. The gray-scale range is indicated by the step
wedge at the right side of the image.
\label{f3}}
\end{figure}

\clearpage
\begin{figure}
\epsscale{1}
\plotone{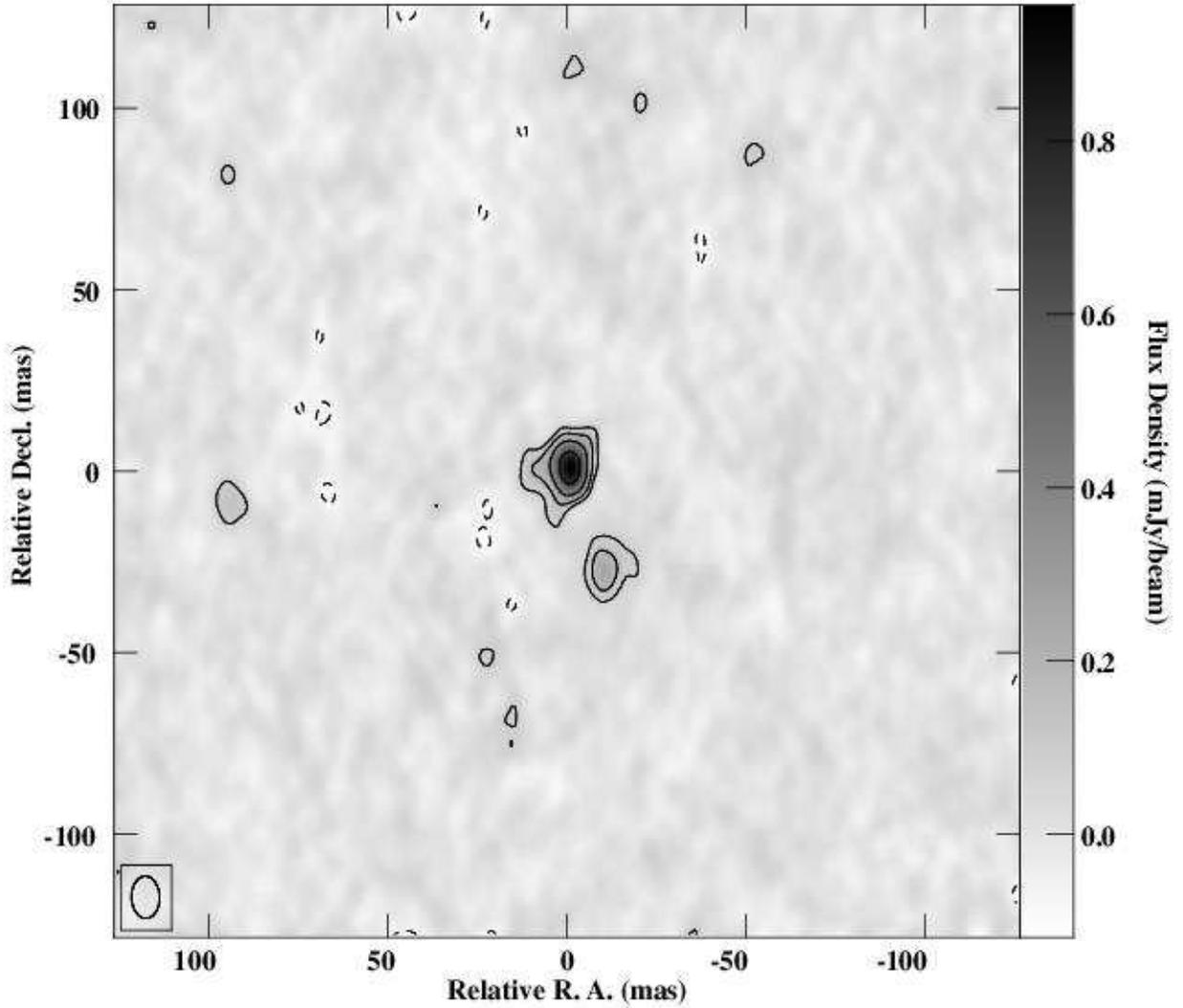}
\figcaption{Naturally weighted VLBA continuum
image of the $z=6.12$ QSO J1427+3312 at 1.4~GHz and $11.6 \times
7.8$~mas resolution (P.~A.=$2^{\circ}$).  The peak flux density is
953~$\mu$Jy~beam$^{-1}$, and the contour levels are at $-3$, 3, 6, 12, 24
times the rms noise level, which is
28~$\mu$Jy~beam$^{-1}$. The gray-scale range is indicated by the
step wedge at the right side of the image.
The reference point (0, 0)
is $\alpha(\rm{J2000.0})= 14^{\rm h}27^{\rm m}38\rlap{.}^{\rm s}5858$,
$\delta(\rm{J2000.0})=+33^{\circ}12^{'}41\rlap{.}^{''}927$.
\label{f4}}
\end{figure}

\clearpage

\begin{deluxetable}{lc}
\tablenum{1}
\tablecolumns{6}\tablewidth{0pc}
\tablecaption{P{\footnotesize ARAMETERS} {\footnotesize OF THE} VLA
O{\footnotesize BSERVATIONS} {\footnotesize OF} J1427+3312}
\tablehead{\colhead{Parameters} & \colhead{Values}}
\startdata
Observing Date \dotfill  & 2007 July 9 \\
Total observing time (hr)\dotfill  & 6 \\
Primary flux calibrator\dotfill & 3C286 \\
Phase calibrator\dotfill  & J1416+3444 \\
Frequency (GHz)\dotfill  &  8.4 \\
Total bandwidth (MHz)\dotfill   & 100 \\
Theoretical noise level ($\mu$Jy~beam$^-1$)\dotfill & 9\tablenotemark{a}\\
Image R.M.S. noise level ($\mu$Jy~beam$^-1$)\dotfill & 12\\
\enddata
\tablenotetext{a}{Assumes natural weighting.}
\end{deluxetable}

\clearpage
\begin{deluxetable}{lc}
\tablenum{2}
\tablecolumns{6}\tablewidth{0pc}
\tablecaption{P{\footnotesize ARAMETERS} {\footnotesize OF THE} VLBA
O{\footnotesize BSERVATIONS} {\footnotesize OF} J1427+3312}
\tablehead{\colhead{Parameters} & \colhead{Values}}
\startdata
Observing Dates \dotfill  & 2007 June 11 \& 12 \\
Total observing time (hr)\dotfill  & 12 \\
Phase calibrator\dotfill  & J1422+3223 \\
Phase-referencing cycle time (min)\dotfill  &  4 \\
Frequency (GHz)\dotfill  &  1.4 \\
Total bandwidth (MHz)\dotfill   & 64\\
Theoretical noise level ($\mu$Jy~beam$^-1$)\dotfill & 24\tablenotemark{a}\\
Image R.M.S. noise level ($\mu$Jy~beam$^-1$) & 28\\
\enddata
\tablenotetext{a}{Assumes natural weighting.}
\end{deluxetable}

\clearpage
\begin{deluxetable}{ccccccccc}
\rotate
\tablenum{3}
\tablecolumns{8}
\tablewidth{0pc}
\tablecaption{G{\footnotesize AUSIAN} F{\footnotesize ITTING} P{\footnotesize ARAMETERS}
{\footnotesize OF THE} C{\footnotesize ONTINUUM} F{\footnotesize EATURES} {\footnotesize IN}
F{\footnotesize IGURE} 4}
\tablehead{
\colhead{Source}
& \colhead{R.~A.~(J2000)}
& \colhead{Decl.~(J2000)}
& \colhead{Relative Position\tablenotemark{a}}
& \colhead{Peak\tablenotemark{b}}
& \colhead{Total}
& \colhead{Deconvolved Size\tablenotemark{c}}
& \colhead{P.A.}
& \colhead{$T_{\rm b} \times 10^{7}$ } \\
\colhead{}
& \colhead{}
& \colhead{}
& \colhead{(mas)}
& \colhead{(mJy~beam$^{-1}$)}
& \colhead{(mJy)}
& \colhead{(mas)}
& \colhead{($^{\circ}$)}
& \colhead{(K)} \\
\colhead{(1)}
& \colhead{(2)}
& \colhead{(3)}
& \colhead{(4)}
& \colhead{(5)}
& \colhead{(6)}
& \colhead{(7)}
& \colhead{(8)}
& \colhead{(9)}}
\startdata
1\dotfill 
& $14^{\rm h} 27^{\rm m} 38\rlap{.}^{\rm s} 5857$
& $+33^{\circ} 12' 41\rlap{.}'' 928$
& 0, 0
& $0.958 \pm 0.028$ 
& $1.089 \pm 0.053$ 
& $4.5~\times 2.7 $
& 173
& 38.6 \\
2\dotfill 
& $14^{\rm h} 27^{\rm m} 38\rlap{.}^{\rm s} 5850$
& $+33^{\circ} 12' 41\rlap{.}'' 899$
& 8.8W, 29S
& $0.262 \pm 0.028$
& $0.343 \pm 0.058$
& $7.3~\times 3.8 $ 
& 175  
& 5.3 \\
3\dotfill 
& $14^{\rm h} 27^{\rm m} 38\rlap{.}^{\rm s} 5866$
& $+33^{\circ} 12' 41\rlap{.}'' 927$
& 11.3E, 1S
& $0.162 \pm 0.028$
& $0.168 \pm 0.050$
& $2.2 \times 1.5 $ 
& 2 
& 21.9 \\
4\dotfill 
& $14^{\rm h} 27^{\rm m} 38\rlap{.}^{\rm s} 5933$
& $+33^{\circ} 12' 41\rlap{.}'' 918$
& 95.4E, 10S
& $0.143 \pm 0.028$
& $0.178 \pm 0.056$
& $6.0~\times 2.3 $ 
& 56  
& 4.4 \\
\enddata
\tablenotetext{a}{With respect to source 1.}
\tablenotetext{b}{Higher than $5\sigma=140\mu$Jy~beam$^{-1}$.}
\tablenotetext{c}{At half maximum.}
\end{deluxetable}

\end{document}